# Mechanical Force-Driven Charge Redistribution for Hydrogen Release at Ambient Conditions in Transition Metal-Intercalated Bilayer Graphene

Jongdeok Kim[1]†, Vikram Mahamiya[2]†, Massimiliano Di Ventra[3], Hoonkyung Lee[1]*

[1]Advanced Materials Program, Department of Physics, Konkuk University, Seoul 05029, Republic of Korea

[2] Condensed Matter and Statistical Physics Section, The Abdus Salam International Centre for Theoretical Physics, Strada Costiera 11, 34151 Trieste, Italy

[3]Department of Physics, University of California, San Diego, La Jolla, California 92093, USA

E-mail: hkiee3@konkuk.ac.kr

Transition-metal (TM) atom-functionalized nanomaterials are promising candidates for hydrogen storage due to their ability to adsorb multiple hydrogen molecules through Kubas interactions. However, achieving efficient hydrogen desorption at ambient conditions remains a critical challenge for practical use. Here, we present a novel approach to modulate the desorption temperature of hydrogen in TM-intercalated bilayer graphene (BLG) using external mechanical forces. By employing first-principles density functional theory (DFT) and thermodynamic occupancy probability calculations, we demonstrate that adjusting the interlayer distance allows for precise control over the interaction energy of $H_2$, thereby facilitating its desorption at ambient conditions. Complete hydrogen desorption occurs when the interlayer distance is reduced below 4.7 Å, 5.3 Å, and 5.1 Å for Sc-, Ti-, and V-intercalated BLG, respectively. Our findings suggest that external mechanical forces can effectively bring hydrogen occupancy to zero by minimizing charge transfer from the TM $d$-orbitals to $H_2$ antibonding orbitals. Notably, while the total charge transferred from the TM atoms remains nearly constant at varying interlayer distances, its redistribution between the graphene layers and $H_2$ fine-tunes the interaction strength. This approach can be extended to large interlayer distances, as supported by recent experiments on graphene oxide membranes [ACS Nano 12, 9309 (2018)]. Furthermore, recent experimental advances in noble gas and alkali metal intercalation in BLG highlight the potential of this approach to overcome the long-standing challenge of high desorption temperatures in TM-functionalized layered nanomaterials.

† *These authors have equally contributed.*

## I. INTRODUCTION

Hydrogen, characterized by its high energy density of 120 MJ/kg, approximately three times that of gasoline (44 MJ/kg) [1], is recognized as a promising substitute for conventional fossil fuels. However, the realization of carbon-neutral hydrogen-based energy resources necessitates the development of a safe and economically viable hydrogen storage method. Although various metal and metal-nonmetal complex hydrides exhibit high hydrogen uptake, meeting U.S. Department of Energy (DOE) criteria [2], the problem is their sluggish kinetic as well as the high enthalpy of adsorption for $H_2$ (50-70 kJ/mol), which make them inappropriate for practical systems [3–6]. On the contrary, physiosorbed materials, predominately metal-organic frameworks, demonstrate high hydrogen uptake at cryogenic temperatures with a low adsorption enthalpy for $H_2$ ( <10 kJ/mol) [7–9]. These materials suffer from the drawback of substantial desorption below room temperature. To address these limitations, a lot of current research extensively focuses on Kubas interactions [10–12], a non-dissociative form of chemical adsorption involving electron donation and back donation between transition metal (TM) *'d'* orbitals and $H_2$ antibonding orbitals [13–15]. Typically, $H_2$ molecules adsorbed via Kubas binding exhibit adsorption enthalpies exceeding 40 kJ/mol [16]. Saillard and Hoffmann [17] applied crystal orbital overlap population (COOP) analysis and partial density of states (PDOS) calculations to characterize the bonding nature in metal-$H_2$ complexes. Bauschlicher and collaborators [18] employed MP2-level quantum chemical calculations to explore how hydrogen molecules interact with a range of 3d metal ions. Lee et al. [19–21] have developed a general statistical framework to achieve practical hydrogen storage capacity in metal-decorated carbon nanostructures, while effectively preventing metal clustering within the system. Recently, Kubas-type $H_2$ storage materials have been synthesized and gained a lot of scientific interest because of the high $H_2$ uptake and fast kinetics [22–25].

Hamaed et al. [26,27] explored low-coordinated Ti-fragments on mesoporous silica, achieving a maximum storage capacity of up to 21.45 wt% at 100 atm, with a corresponding $H_2$ binding enthalpy of 22.15 kJ/mol, indicative of Kubas-type interactions. Computational studies by Skipper and coworkers [28] on early TM-metal grafted mesoporous silica revealed the presence of Kubas interactions between Ti and $H_2$ molecules, aligning with experimental findings. They have also reported a slight increase in the H-H bond length from 0.75 Å to 0.77 Å – 0.79 Å, which is also a prominent feature of Kubas binding. Additionally, TM-doped porous zeolite and metal-organic frameworks are found to be suitable hydrogen storage systems in experiments due to the Kubas binding between TM and H$_2$ [29–31]. Early TMs (3d/4d series) decorated nanoscale systems are sound hydrogen storage systems [32–37]. Skipper et al. [38] have investigated the H$_2$-storage performance of M (II) (M = Ti, V, Cr) dispersed hydrazine-based materials and compared it with the experimental data. They reported that Mn is not suitable for H$_2$-storage, however, Ti and V dispersed hydrazine can have 7 % and 5 % of H$_2$ uptake, respectively. This study demonstrates that early TM atoms should be chosen for doping or decoration purposes. Hamead and coworkers [39] have also chosen early TM atoms for the functionalization of porous silica in recent experiments. They reported a maximum hydrogen uptake of 3.05 % at 77 K and 85 bar. Although TM-functionalized nanostructures can bind multiple H$_2$ molecules via Kubas interaction, the desorption of these hydrogen molecules typically does not occur under ambient conditions, limiting their practical application. Designing novel materials that not only bind multiple hydrogen molecules but also release them near ambient conditions remains a significant challenge in hydrogen storage research.

Recent experimental studies on noble gas and alkali metal-functionalized bilayer graphene (BLG) under ambient conditions have opened new research directions toward metal-encapsulated two-dimensional van der Waals solids for energy storage [40,41]. Liu et al. [42] demonstrated that multi-layered Ti$_2$CT$_x$, a type of MXene, achieves a hydrogen (H$_2$) storage capacity of up to 8.8 wt% at room temperature and 60 bar H$_2$ pressure. Remarkably, even under ambient conditions, Ti$_2$CT$_x$ retains approximately 4 wt%. The study revealed that Ti atoms in the system exhibit interactions with H$_2$ molecules akin to the Kubas mechanism. Moreover, the nanopump effect, driven by an optimal interlayer spacing of ~6.8 Å, enhances both binding energies and storage capacity. In contrast, at an interlayer distance of 8 Å, the interaction is predominantly physisorption.

Motivated by these findings, we investigated the interaction of hydrogen molecules adsorbed on TM atoms encapsulated within BLG. We propose a novel strategy for regulating the elevated desorption temperatures of TM-functionalized layered nanostructures through the application of mechanical forces. This approach could be instrumental in the design of practical hydrogen storage materials capable of efficiently releasing hydrogen under ambient conditions. A recent experimental study also reported that controlling the interlayer spacing of two-dimensional materials at the angstrom level via external pressure regulation (EPR) could be employed in water purification [43]. In this approach, external pressure was modulated by adjusting screw tension, with graphene oxide membranes compressed using punched steel plates. This effective method holds potential for tuning material properties in energy storage applications by controlling interlayer distances.

## II. COMPUTATIONAL METHOD

Computational calculations are performed employing density functional theory (DFT) using the Vienna Ab Initio Simulation Package (VASP 5.4.4) [44], implementing the projector-augmented-wave (PAW) method [45]. The Perdew−Burke−Ernzerhof (PBE) exchange-correlation functional within the generalized gradient approximation (GGA) framework [46] was employed, while Grimme's DFT-D3 dispersion corrections were incorporated to account for weak van der Waals interactions between hydrogen and the sorbent materials [47]. Spin polarization was explicitly included in all calculations to accurately capture the magnetic effects associated with the transition metal atom. The Brillouin zone was sampled using a Monkhorst–Pack $k$-point mesh of 4×4×1, and a plane-wave energy cutoff of 500 eV was employed. The convergence criteria for total energy and atomic forces were set to $10^{-5}$ eV and 0.05 eV/Å, respectively. A vacuum layer of 20 Å was introduced along the $z$-direction to eliminate spurious interactions between periodic images. All calculations were conducted on a 4×4 supercell of BLG, with a single TM atom embedded to minimize the possibility of metal clustering. Geometry optimizations were performed under constrained conditions, with the interlayer spacing held fixed, thereby excluding the effects of buckling in this study.

## III. RESULTS AND DISCUSSION

We initiated the calculations by functionalizing a monolayer graphene (MLG) with light TMs, specifically scandium (Sc), titanium (Ti), and vanadium (V), which have been experimentally validated as effective sorption centers for Kubas-type interactions [48]. Light TM atoms are known to bind multiple $H_2$ molecules through the Kubas interaction, facilitated by the availability of vacant $d$-orbitals. We observed that up to five, four, and four hydrogen molecules can be effectively adsorbed at the Sc, Ti, and V sites of functionalized MLG, respectively. Notably, in the case of Ti-functionalized graphene, the first $H_2$ molecule dissociates and binds in the form of isolated hydrogen atoms, resulting in a higher binding energy compared to subsequent hydrogen molecules. The dissociation of first $H_2$ molecule on Ti can be attributed to the small overlap between Ti-$d$ orbitals and $H_2$ antibonding orbitals, which has been observed for such light TMs dopant [24]. Interestingly, multiple $H_2$ molecules are bound in molecular form on Ti atom in its ground state configuration.

**Figure 1 (a)-(d)** presents the optimized geometries of Ti-functionalized MLG with varying numbers of adsorbed hydrogen molecules (1st $H_2$ gets dissociated), ranging from one to four. The corresponding binding energies for $H_2$ molecules adsorbed on Sc, Ti, and V-functionalized MLG nanostructures are shown in **Figure 1 (e)**. The average binding energy $\varepsilon_b$ for all the adsorbed $H_2$ on a TM site was calculated using the following equation:

$$\varepsilon_b = \frac{E_{MLG+TM+nH_2} - (E_{TM+MLG} + nE_{H_2})}{n} \qquad (1)$$

where $n$ represents the number of adsorbed hydrogen molecules. $E_{MLG+TM}$ and $E_{MLG+TM+nH_2}$ are the total energies of the TM-functionalized graphene monolayer and the system with $n$ adsorbed $H_2$ molecules, respectively, and $E_{H_2}$ is the total energy of an isolated hydrogen molecule. All calculated binding energies are negative, indicating that the adsorption of hydrogen molecules is energetically favorable. **Figure 1 (f)** depicts the magnitude of the adsorption energies, with a gray-shaded region highlighting the range of 0.2–0.4 eV, which corresponds to the DOE-criterion for reversible hydrogen storage. Notably, full desorption of adsorbed hydrogen molecule can occur near ambient temperature when hydrogen binding energy is ~0.2 eV.

The binding energies for all $H_2$ molecules adsorbed on Sc, Ti, and V-functionalized MLG exceeded the DOE's range for reversibility, except for the first $H_2$ molecule adsorbed on Sc-functionalized graphene. The maximum hydrogen binding energies were calculated as 0.54, 0.76, and 0.91 eV for Sc, Ti, and V sites, respectively. These values significantly surpass the energy range needed for reversible hydrogen storage near room temperature, suggesting that $H_2$ release from TM-functionalized MLG nanostructures would not readily occur under ambient conditions. To investigate the hydrogen adsorption behavior on TM-functionalized MLGs, we used the grand canonical partition function to calculate the occupation number ($f$) at each metal site as a function of temperature and pressure, using equation (2) [20]:

$$f = \frac{\sum_{n=0} n e^{n[\mu(p,T)-\varepsilon_b]/k_\beta T}}{\sum_{n=0} e^{n[\mu(p,T)-\varepsilon_b]/k_\beta T}} \qquad (2)$$

where $\varepsilon_b$ is the adsorption energy per $H_2$ molecule, $n$ is the number of adsorbed hydrogen molecules, $\mu(p,T)$ represents the chemical potential of hydrogen gas at a given pressure $p$ and temperature $T$, and $k_\beta$ denotes the Boltzmann constant. Equation (2) is fitted to experimental values of the H$_2$ chemical potential as a function of temperature ($T$) and pressure ($p$).

**Figure 2 (a)-(d)** illustrate the occupancy of hydrogen molecules on each TM-functionalized MLG nanostructure as a function of temperature, under constant hydrogen pressures (1, 10, 20, and 100 bar). As expected, occupancy decreases with increasing temperature and increases with higher pressure. This reflects the well-known trend of achieving higher hydrogen storage capacity at lower temperatures and higher pressures. For Sc-functionalized MLG, full desorption occurs at approximately 550 K at 1 bar, while for Ti- and V-functionalized graphene, desorption occurs at around 700 K and 800 K, respectively. Under high-pressure conditions, the desorption temperature increases, reaching up to 800 K for all metal-functionalized systems at 100 bars. These results arise due to the stronger binding energies of H$_2$ molecules at high pressure enabling higher desorption temperatures for hydrogen release. In our previous work [20], we recommended a practical adsorption and desorption conditions for hydrogen storage which are 298 K and 30 bar for H$_2$ adsorption, and a desorption temperature and pressure of 373 K and 2 bar, respectively. Bhatia and Myers [49] proposed a hydrogen delivery pressure of 1.5 bar at 298 K for carbon-based sorbents. Importantly, desorption temperatures should not exceed 373 K to avoid safety concerns and minimize the additional energy required for hydrogen release. Although TM-functionalized MLG nanostructures can bind multiple hydrogen molecules via Kubas interaction, desorption does not readily occur near ambient conditions, rendering these systems inefficient under practical operating conditions.

Recently, noble metal encapsulated BLG nanostructures have been experimentally observed near ambient temperature [40]. Additionally, the successful experimental intercalation of various alkali metal atoms (Li, K, Rb, Cs) into graphene layers [41,50], further motivates the exploration of TM-encapsulated BLG nanostructures. Our study aims to facilitate hydrogen release under near-ambient conditions, potentially advancing the design of efficient hydrogen storage materials. We investigated the hydrogen storage properties of various TM atoms (Sc, Ti, V) intercalated BLG nanostructures. **Figure 3** show the geometries of hydrogen molecules adsorbed on different TM-intercalated BLG nanostructures with varying interlayer spacing (~4–6 Å). The interlayer distance of the BLG is varied in increments of 0.3 Å, demonstrating how the positions and interaction of the adsorbed hydrogen molecules change at varying interlayer distances. In the case of titanium, we observed the adsorption of a single hydrogen molecule in a dissociated form. To investigate H$_2$ adsorption due to quasi-molecular Kubas binding, we performed additional optimization calculations for a system with two hydrogen molecules, varying the interlayer distance. As shown in **Figure 3 (b)**, the two vertically adsorbed $H_2$ molecules rotate into a horizontal position at an interlayer distance of ~4.8 Å. In Ti-intercalated systems with interlayer distances of ~4.8 Å or less, the $H_2$ molecules adsorb in a horizontal configuration, leading to a lower total energy and a more stable structure. We observed that the orientation of the adsorbed hydrogen molecules shifts from a vertical bond (VB) facing to a horizontal bond (HB) facing for all TMs at reduced interlayer distances which increases the interactions between TM and graphene layers. However, this change is relatively small in the case of V-intercalated BLG. **Figure 4 (a)-(c)** present the calculated binding energies per $H_2$ molecule on various TM-intercalated BLG nanostructures as a function of interlayer distance. The results indicate that the average binding energy of adsorbed $H_2$ molecules initially increases linearly with increasing interlayer distance, eventually reaching a plateau at an interlayer distance greater than ~6 Å. The maximum binding energy for $H_2$ molecules adsorbed on intercalated TM atoms was found to be 0.52 eV, 0.68 eV, and 0.70 eV for Sc, Ti, and V, respectively, at larger interlayer distances (d > 6 Å). The grey-shaded region in **Figure 4 (a)-(c)** indicates the reversible adsorption energy range. Hydrogen adsorption occurs reversibly when the interlayer distances are within the ranges of 4.7–5.2 Å, 5.3–5.5 Å, and 5.1–5.3 Å for Sc, Ti, and V atoms, respectively. This implies that complete desorption of $H_2$ molecules can be achieved if the interlayer distance is reduced below 4.7 Å, 5.3 Å, and 5.1 Å for Sc-, Ti-, and V-intercalated BLG nanostructures under ambient conditions. Therefore, by systematically controlling the interlayer distances in layered nanostructures, the hydrogen adsorption energy can be tuned to achieve values within the range required for reversible adsorption or even lower, facilitating hydrogen desorption under ambient conditions. Since the adsorbed hydrogen molecules bind vertically to the intercalated metal centers, the EPR strategy can be applied for interlayer distances up to 10 Å, as demonstrated in recent experiments on graphene oxide membranes [43]. It is important to note that real systems often exhibit local structural variations, such as partial staging, local expansion, or defects, especially upon metal intercalation. These features may affect the hydrogen binding characteristics. However, a detailed investigation of such effects lies beyond the scope of the present study.

Next, we determined the mechanical forces and external pressure required to maintain specific interlayer spacings in our calculations. To achieve this, we calculated the total energy variation of hydrogen-adsorbed TM-intercalated BLG nanostructures as a function of interlayer distance. The corresponding vertical forces and applied external pressure as a functional of interlayer spacing are plotted in **Figure 4 (d)–(f)**. Our results show that the equilibrium interlayer distances, where the vertical forces acting on $H_2$-adsorbed Sc-, Ti-, and V-intercalated BLG nanostructures become zero are 4.1 Å, 5.9 Å, and 5.3 Å, respectively. Positive and negative values of the vertical forces correspond to repulsive and attractive interactions between the graphene layers. For Sc-intercalated BLG, increasing the interlayer distance beyond the equilibrium position is required to shift the $H_2$ adsorption energy into the range suitable for reversible Kubas-type hydrogen storage. Conversely, for Ti-intercalated BLG, a reduction in the interlayer spacing is necessary. As a result, tensile forces are required for Sc, while external compressive forces are needed for Ti to achieve reversible $H_2$ storage at ambient conditions. Interestingly, for V-intercalated BLG, the vertical forces are zero at an interlayer spacing of 5.3 Å, which lies within the reversible hydrogen adsorption energy window. **Figures 4 (d)–(f)** illustrate that tuning the hydrogen adsorption energy into the reversible storage regime for Sc- and Ti-intercalated BLG systems requires the application of external pressures in the range of ~1–3 GPa. Interestingly, the V-intercalated BLG system achieves this tuning at pressures nearly ten times lower, making it a more promising candidate for low-pressure hydrogen storage. This observation aligns well with the experimental findings of Vasu et al. [51], who reported that trapped molecules between graphene layers can experience local pressures on the order of ~1 GPa supporting the feasibility of pressure-modulated hydrogen control in such layered systems. Thus, by applying a controlled external mechanical force of appropriate magnitude and direction, it is possible to maintain BLG nanostructures at desired interlayer distances.

Recent work by Lin et al. [41] reported an interlayer distance of 5.95 Å for Cs-intercalated BLG, while Liu et al. experimentally observed an interlayer spacing of 6.8 Å for $Ti_2CT_x$ MXene in the context of Kubas-type hydrogen binding [42]. These findings are consistent with our results, where we observe that the Kubas interaction is the dominant mechanism governing $H_2$ binding when the interlayer spacing is within the range of 6–7 Å, as illustrated in **Figure 4 (a)-(c)**. Additionally, we found that the $H_2$ binding energy in BLG without TM-intercalation can be tuned to fall within the reversible adsorption window by maintaining an interlayer distance between 4.6 and 4.8 Å. Desorption of adsorbed $H_2$ occurs when the interlayer distance exceeds 4.8 Å under ambient conditions. This behavior is attributed to enhanced intrinsic polarization resulting from charge redistribution within the graphene layers as the interlayer spacing decreases, a phenomenon analogous to the application of in-plane compressive strain in MLG to enhance binding strength [52]. In TM-intercalated BLG nanostructures, the release of $H_2$ molecules at ambient conditions can be attributed to direction-dependent forces, with both the magnitude and direction of these forces being determined for the specific TM atoms intercalated. While the choice of exchange-correlation functional or dispersion correction may slightly influence the calculated hydrogen binding energy, a characteristic interlayer spacing window suitable for reversible hydrogen adsorption can still be identified for each transition metal atom, regardless of the specific functional used. Similarly, the inclusion of zero-point energy (ZPE) can account for up to 25% of the static adsorption energy in metal-decorated graphene systems [20], and thus can significantly reduce the effective hydrogen binding energy. This reduction may broaden the interlayer spacing window favorable for reversible hydrogen adsorption.

To investigate the origin of this direction-dependent, force-driven hydrogen release at ambient conditions, we calculated the Bader charge on the TMs, BLG, and $H_2$ molecules, as shown in **Figure 4 (g)-(i)**. We found that the charge depletion on the TM atoms remains relatively constant with varying interlayer spacing, but the redistribution of charge between the graphene layers and $H_2$ molecules changes, effectively tuning the host-adsorbent interactions. Among the TMs considered, Sc exhibited the highest Bader charge depletion. Notably, most of the charge lost by Sc is transferred to the graphene layers for interlayer spacings between 4 and 6 Å, with the remaining charge accumulating on the adsorbed $H_2$. The atomic radius of Sc is 0.15 Å larger than that of Ti, and 0.25 Å larger than V, resulting in a stronger interaction between Sc's valence shell *d*-electrons and the C-*2p* orbitals of graphene layers. This increased charge transfer to the graphene layers leaves a moderate amount of charge for transfer to the $H_2$ molecules, bringing the $H_2$ binding energy near the reversible adsorption range (0.1 - 0.5 eV/$H_2$) for all considered interlayer spacings. In the case of Ti, significant charge (~1.5 e) is transferred from the Ti valence shells to the graphene layers at an interlayer spacing of ~5 Å. However, as the interlayer distance increases, the interaction between Ti and the graphene layers weakens, leading to greater charge transfer to the $H_2$ molecules at distances greater than 5 Å, resulting in a comparatively high $H_2$ binding energy (~ 0.7 - 0.8 eV/$H_2$). Vanadium exhibits a similar trend, with decreasing charge transfer from its valence shells to the graphene layers at larger interlayer spacings, leading to increased charge gain by $H_2$. **Figure 5 (a)-(c)** present the charge density difference $\left(\Delta\rho = \rho_{total} - \rho_{BLG+TM} - \rho_{H_2}\right)$ in TM-intercalated systems, with all the considered interlayer spacings. These results confirm the presence of Kubas-type interactions, characterized by electron exchange between the TM-atoms and $H_2$. At interlayer distances exceeding 6 Å, the interaction between the TM-atoms and the graphene layers becomes minimal and nearly invariant. Consequently, the hydrogen binding energy remains relatively high and constant beyond this threshold. In contrast, as the interlayer spacing decreases, a notable charge depletion is observed in the region between the upper graphene layer and the hydrogenated TM atoms. This indicates electron migration from the TM centers toward the graphene, particularly the upper layer. Such increased charge transfer to the graphene weakens the interaction between the TM-*d* orbitals and the antibonding orbitals of $H_2$, thus facilitating the release of $H_2$ molecules under ambient conditions. To further probe these interactions, **Figure 6 (a)-(f)** present the projected density of states (PDOS) for the Ti-*4s*, Ti-*3d*, and $H_2$-*1s* orbitals at interlayer spacings of d = 4.7 Å, 5.3 Å, and 5.9 Å. At the smallest spacing (4.7 Å), the Ti-*3d* and $H_2$-*1s* orbitals exhibit negligible overlap, as most of the electron density depleted from Ti is transferred to the graphene layers. However, at increased spacings of 5.3 Å and 5.9 Å, a clear hybridization between the Ti-3*d* and $H_2$-*1s* states emerges near the Fermi level, indicative of a stronger orbital interaction and a more pronounced Kubas interaction. These findings underscore the critical role of interlayer spacing in tuning the electronic interactions and binding strength of $H_2$ in TM-intercalated BLG systems.

Metal clustering is a well-known challenge in TM-functionalized nanostructures, as it can significantly reduce hydrogen uptake and hinder practical applications. To evaluate the likelihood of TM atom clustering in BLG, we calculated the diffusion energy barrier for a single TM atom migrating from one stable site to the nearest equivalent site using the climbing-image nudged elastic band (CI-NEB) method, as illustrated in **Figures 7(a)-(c).** For these calculations, six intermediate images were considered along the diffusion path between two adjacent hollow sites located on neighboring benzene rings, and the corresponding energy profile is presented in **Figure 7(d)**. In our model system, one TM atom is embedded within a 4×4 supercell of BLG. In this configuration, the effective energy barrier for TM dimer formation can be up to four times the calculated single-hop diffusion barrier shown in **Figure 7(d)**. Among the studied metals, vanadium exhibits the lowest single-hop diffusion barrier of 0.47 eV, corresponding to an effective barrier in the range 0.47-1.88 eV for dimer formation in the 4 × 4 supercell. This range is substantially higher than the thermal energy at room temperature (~25 meV), indicating that TM atom migration and subsequent clustering are energetically unfavorable under ambient conditions. Therefore, metal clustering is unlikely in TM-embedded BLG systems, ensuring their structural stability and suitability for hydrogen storage applications.

To further assess the hydrogen storage performance, we carried out thermodynamic occupancy calculations of $H_2$ as a function of interlayer spacing under varying conditions.

These calculations were performed at $H_2$ pressures of 1, 10, and 20 bar, and temperatures of 270 K, 300 K, 350 K, and 400 K. As illustrated in **Figure 8 (a)-(c)**. For a fixed interlayer spacing, increasing the temperature leads to a reduction in hydrogen occupancy, indicating progressive desorption of $H_2$ molecules. Notably, at 1 bar and near-ambient or lower temperatures, the hydrogen occupancy drops to zero at interlayer distances of approximately 4.7 Å, 5.3 Å, and 5.1 Å for Sc, Ti, and V intercalated BLG systems, respectively. These findings suggest that complete hydrogen desorption can be realized under practical conditions through precise control of the interlayer spacing in layered nanomaterials, offering a viable strategy for reversible hydrogen storage.

## IV. CONCLUSION

We have investigated hydrogen adsorption in transition metal (TM: Sc, Ti, V) intercalated BLG systems by systematically tuning the interlayer spacing through the application of external mechanical forces. Our results reveal that at lower interlayer spacings, most of the charge transferred from the TM atom accumulates on the upper graphene layer, with a smaller fraction localized on the $H_2$ molecules. In contrast, at larger spacings, the TM-to-graphene charge transfer becomes negligible, and the charge is predominantly transferred to the $H_2$ molecules, resulting in stronger hydrogen adsorption. The mechanical pressure required to achieve these variations in interlayer spacing was explicitly calculated. Additionally, the electronic density of states indicates that at small interlayer distances, the overlap between TM-$d$ and $H_2$-1s orbitals is minimal. However, as the spacing increases, a clear hybridization emerges, consistent with the formation of reversible Kubas-type interactions. Thermodynamic occupancy calculations further demonstrate that at 1 bar and near-ambient or lower temperatures, hydrogen occupancy drops to zero at interlayer distances of approximately 4.7 Å, 5.3 Å, and 5.1 Å for Sc-, Ti-, and V-intercalated systems, respectively, implying that complete hydrogen desorption can be achieved under practical operating conditions. These findings highlight the critical role of charge redistribution between the graphene host layers and $H_2$ molecules in governing the binding mechanism. The ability to regulate this charge transfer by adjusting the interlayer spacing provides a viable strategy for tuning hydrogen interaction energies and enabling controlled desorption. Overall, our study presents a promising route for realizing efficient and reversible hydrogen storage in layered nanostructures through external mechanical modulation, in alignment with recent experimental findings on metal-intercalated bilayer graphene.

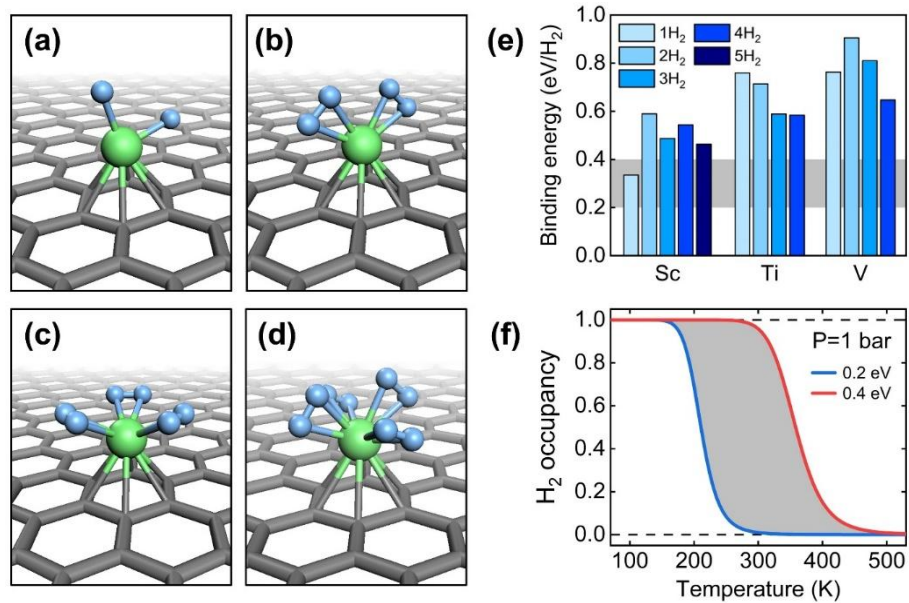

**Figure 1.** Optimized structures of titanium-functionalized monolayer graphene with (a) dissociated $H_2$ (b) $2H_2$ (c) $3H_2$ (d) $4H_2$. Here, green, and cyan balls represent Ti and H atoms, respectively. (e) Binding energy of adsorbed $H_2$ molecules on various TM-functionalized monolayer graphene where grey area indicates the range of binding energies required for reversible hydrogen storage. (f) Temperature-dependent thermodynamic occupancy of $H_2$ molecules for reversible hydrogen adsorption energy window.

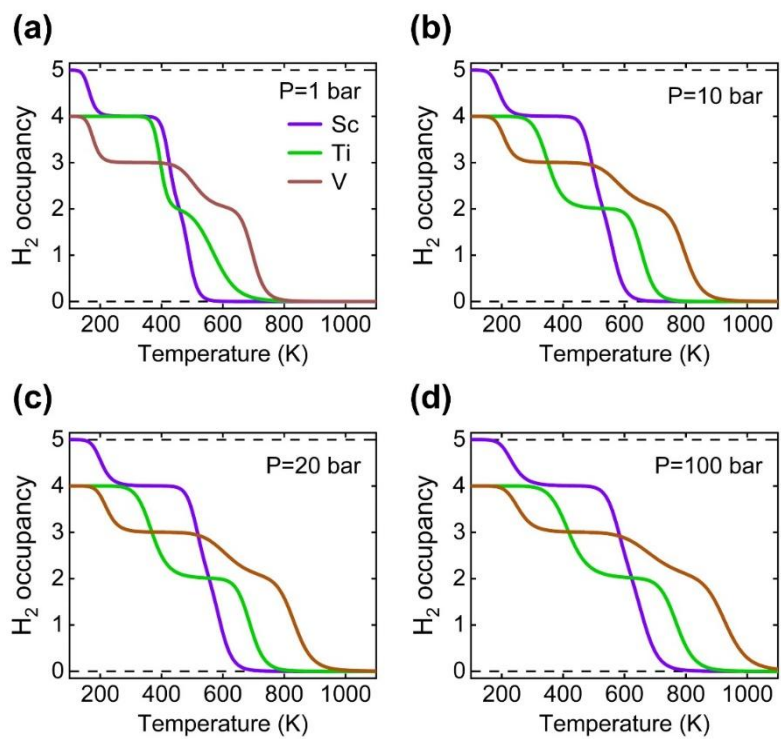

**Figure 2.** Thermodynamic occupancy of $H_2$ molecules adsorbed on Sc, Ti, and V atoms-functionalized monolayer graphene at (a) 1, (b) 10, (c) 20, (d) 100 bars of $H_2$ pressure. Here, purple, green, and brown lines indicate Sc, Ti, and V-functionalized monolayer graphene nanostructures, respectively.

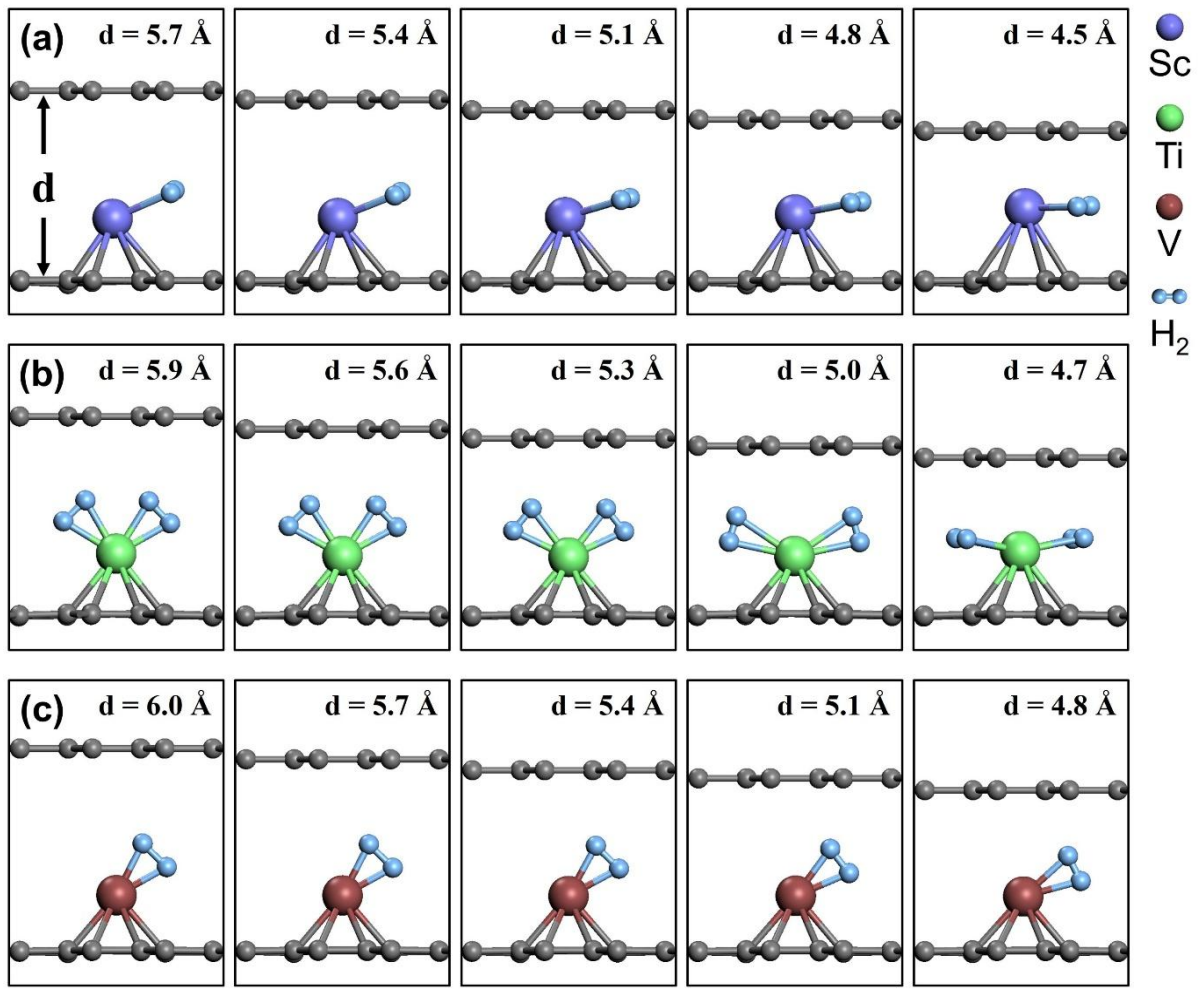

**Figure 3.** Optimized geometries of $H_2$ molecules adsorbed on various TM-intercalated bilayer graphene nanostructures placed at different interlayer distances, (a) $H_2$ adsorbed on Sc-intercalated bilayer graphene, (b) $2H_2$ adsorbed on Ti-intercalated bilayer graphene, and (c) $H_2$ adsorbed on V-intercalated bilayer graphene. Grey balls represent carbon atoms.

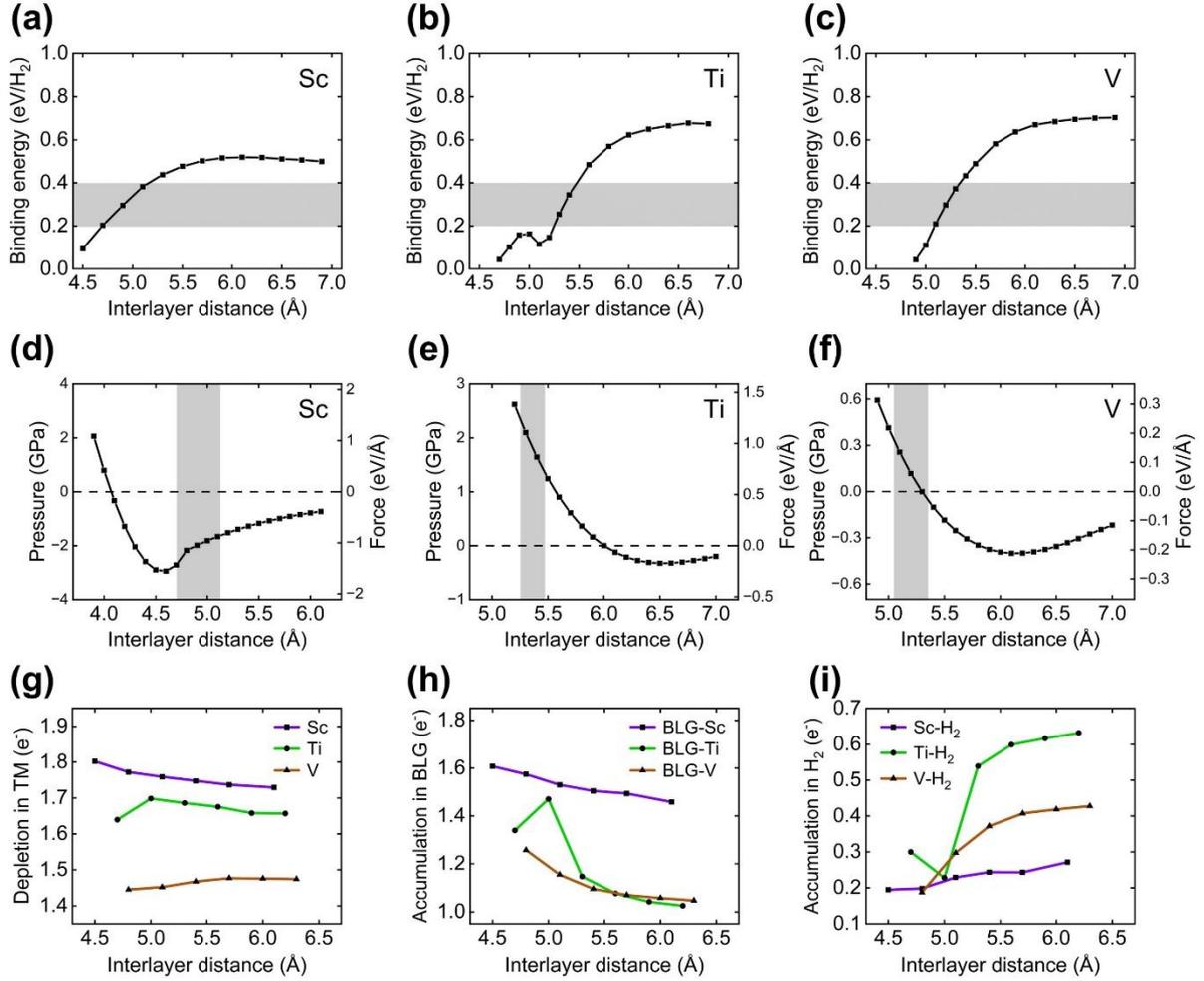

**Figure 4.** (a)-(c) Binding energy of $H_2$ molecules on light TMs (Sc, Ti, V)-intercalated bilayer graphene as a function of interlayer distance. (d)-(f) Calculated vertical mechanical force and external pressure acting on TM-intercalated bilayer graphene as a function of interlayer distance. Positive and negative values of force represent repulsion and attraction between the layers, respectively. Grey area indicates the reversible adsorption of $H_2$ molecules under ambient conditions. (g)-(i) Site specific Bader charge values for Sc, Ti, and V-intercalated bilayer graphene as a function of interlayer distance.

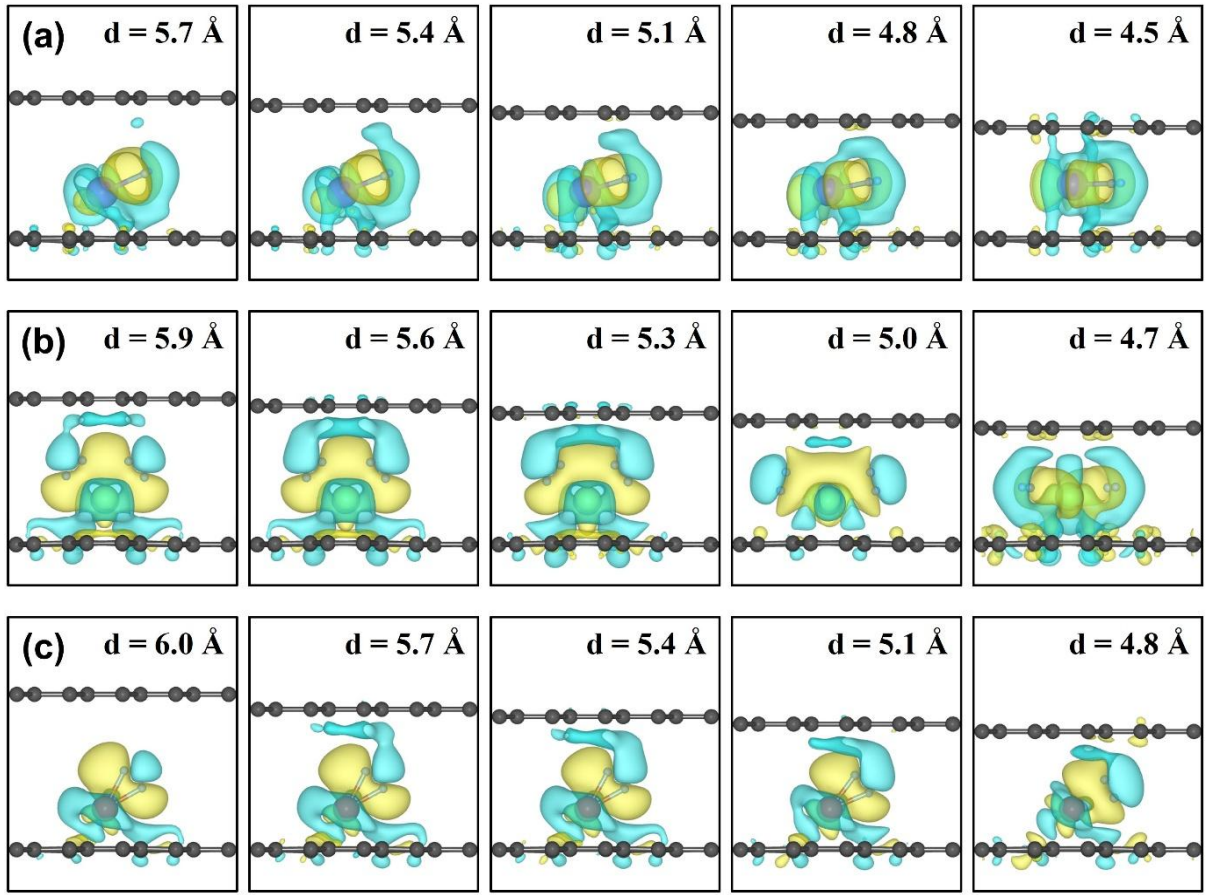

**Figure 5.** Charge density difference plots of (a) Sc, (b) Ti and (c) V-intercalated bilayer graphene nanostructures at different interlayer distances. The yellow and cyan regions represent negative charge accumulation and depletion, respectively. The isosurface value of the charge density was set to $8.0 \times 10^{-4}$ e/Å$^3$.

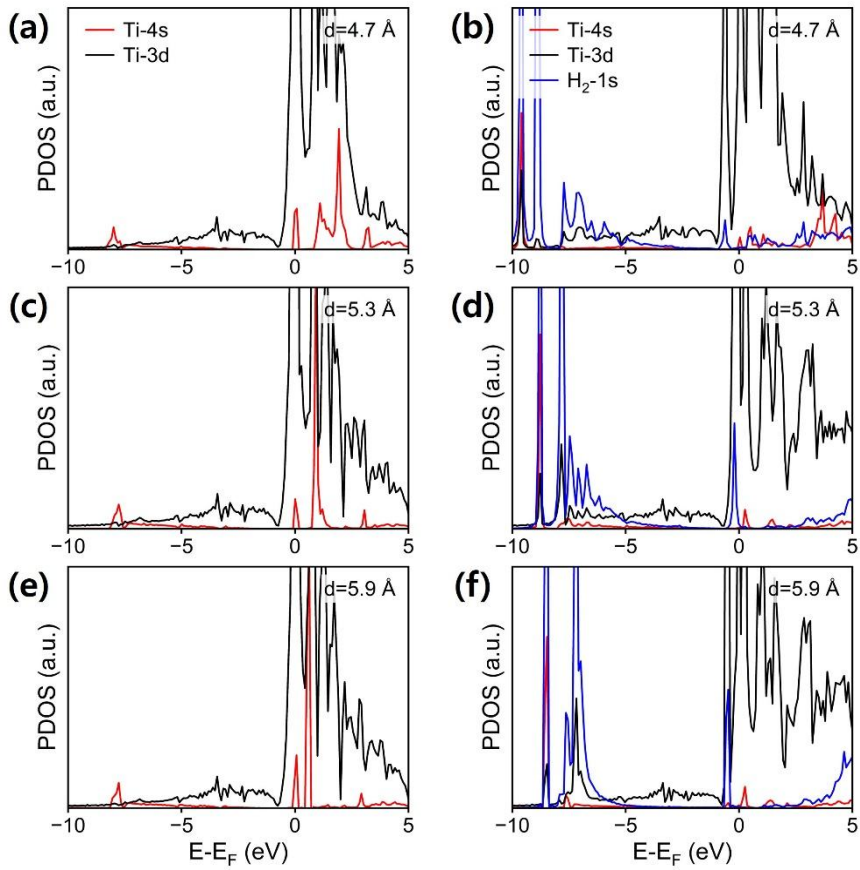

**Figure 6.** Projected density of states (PDOS) for Ti-intercalated bilayer graphene at varying interlayer spacings: (a, c, e) PDOS of Ti 4s and 3d orbitals in BLG + Ti at $d$ = 4.7 Å, 5.3 Å, and 5.9 Å, respectively; (b, d, f) PDOS of Ti 4s, Ti 3d, and $H_2$ 1s orbitals in BLG + Ti + $H_2$ at the same corresponding spacings. The Fermi energy ($E_F$) is set to 0 eV in all plots.

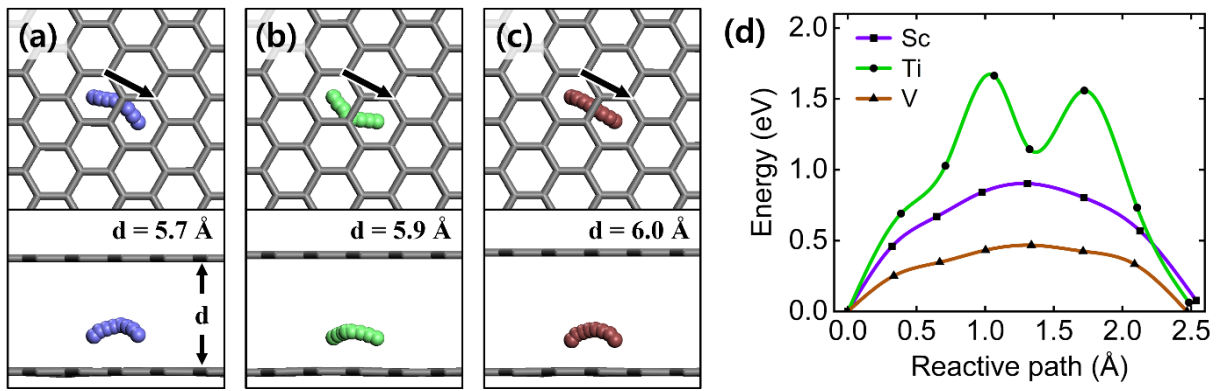

**Figure 7.** Minimum energy diffusion paths for TM atoms embedded in bilayer graphene, calculated using the climbing-image nudged elastic band (CI-NEB) method: (a) Sc at an interlayer spacing of $d$ = 5.7 Å, (b) Ti at $d$ = 5.9 Å, and (c) V at $d$ = 6.0 Å. (d) Corresponding energy profiles along the reaction coordinate, showing the diffusion barriers for TM atom migration between adjacent hollow sites in BLG.

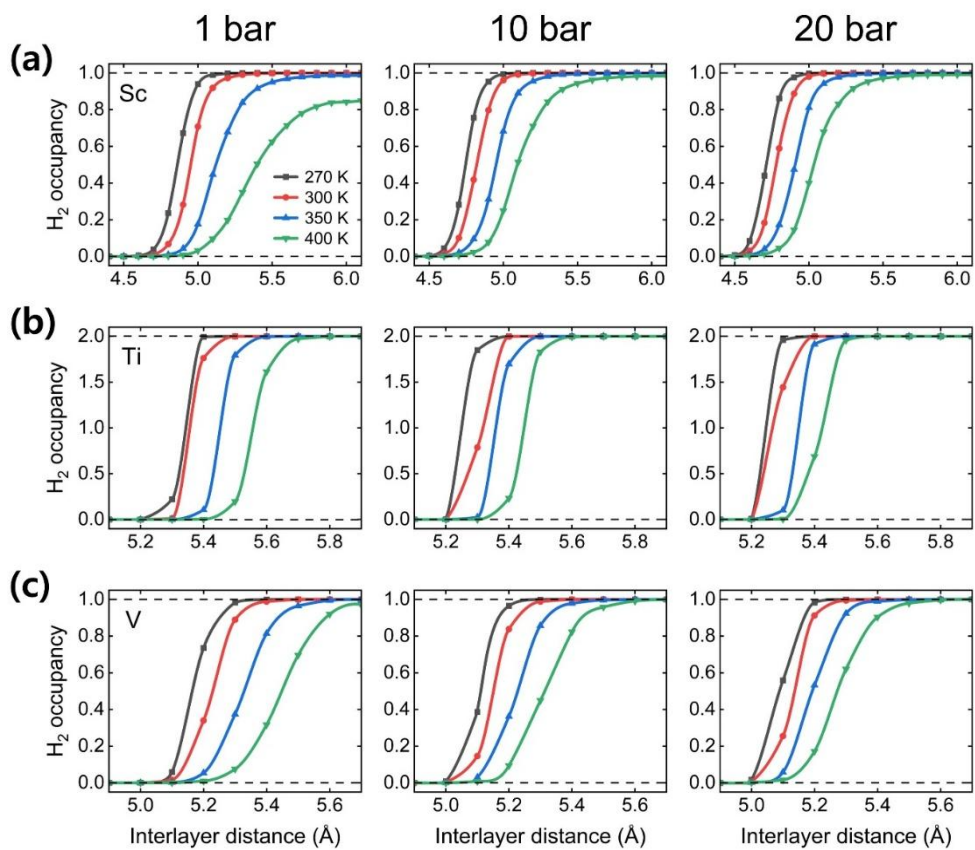

**Figure 8.** Thermodynamic occupancy of $H_2$ molecules on (a) Sc, (b) Ti, and (c) V-intercalated bilayer graphene as a function of interlayer distance at 1, 10, and 20 bar of $H_2$ pressure. Black, red, blue, and green lines indicate the temperature values of 270, 300, 350, and 400 K, respectively.

**Acknowledgements**

This project was supported by Konkuk University, South Korea in 2022.

**Author Contributions**

The manuscript was written through the contributions of all authors. All authors have given approval to the final version of the manuscript. Jongdeok Kim and Vikram Mahamiya equally contributed to this work. Correspondence should be addressed to H. Lee* (hkiee3@konkuk.ac.kr).

*Corresponding author email: Hoonkyung Lee: hkiee3@konkuk.ac.kr*

*Contact author email: Vikram Mahamiya: vmahamiy@ictp.it , Jongdeok Kim: loving623@naver.com*